\documentclass[12pt,preprint]{aastex}

\begin{document}


\title{{\it STEREO} Direct Imaging of a CME-driven Shock to 0.5\,AU}

\author{Shane A. Maloney and Peter T. Gallagher} \affil{School of Physics, Trinity College Dublin, Dublin 2, Ireland}

\begin{abstract}
\label{abs}

Fast coronal mass ejections (CMEs) generate standing or bow shocks as they propagate through the corona and solar wind. Although CME shocks have previously been detected indirectly via their emission at radio frequencies, direct imaging has remained elusive due to their low contrast at optical wavelengths. Here we report the first images of a CME-driven shock as it propagates through interplanetary space from 8\,R\,$_{\odot}$ to 120\,R\,$_{\odot}$ (0.5\,AU), using observations from the {\it STEREO} Heliospheric Imager (HI). The CME was measured to have a velocity of $\sim$1000\,km\,s$^{-1}$ and a Mach number of 4.1$\pm$1.2, while the shock front stand-off distance ($\Delta$) was found to increase linearly to $\sim$20~$R_\odot$ at 0.5~AU. The normalised standoff distance ($\Delta/D_{O}$) showed reasonable agreement with semi-empirical relations, where $D_O$ is the CME radius. However, when normalised using the radius of curvature,  $\Delta/R_{O}$ did not agree well with theory, implying that $R_{O}$ was under-estimated by a factor of $\approx$3--8. This is most likely due to the difficulty in estimating the larger radius of curvature along the CME axis from the observations, which provide only a cross-sectional view of the CME.\\
\end{abstract}



\section{Introduction} 
\label{s_intro}
Bow shocks occur when a blunt object moves relative to a medium at supersonic speeds \citep{rathakrishnan:2010applied}. These shocks are formed across many scales and in different conditions; from astrophysical shocks such as planetary bow shocks \citep{Slavin:1981p11401}, or the shock at the edge of the Heliosphere \citep{vanBuren:1995p2914}, to shocks generated by the reentry of the Apollo mission capsules \citep{Glass:1977p269}. Coronal mass ejections (CMEs) which travel faster than the local fast magnetosonic velocity (with respect to the solar wind velocity) produce such standing shocks in the frame of the CME \citep{Stewart:1974p9160, Stewart:1974p9161}. Interplanetary (IP) CME-driven shocks have previously been detected in radio observations as Type II bursts and using {\it in-situ} measurements. Direct imaging of shocks, on the other hand, has remained elusive, primarily due their low contrast {\citep{Vourlidas:2009p8783, Gopalswamy:2008p4948}.

\begin{figure}[!ht]
	\epsscale{0.5}
	\plotone{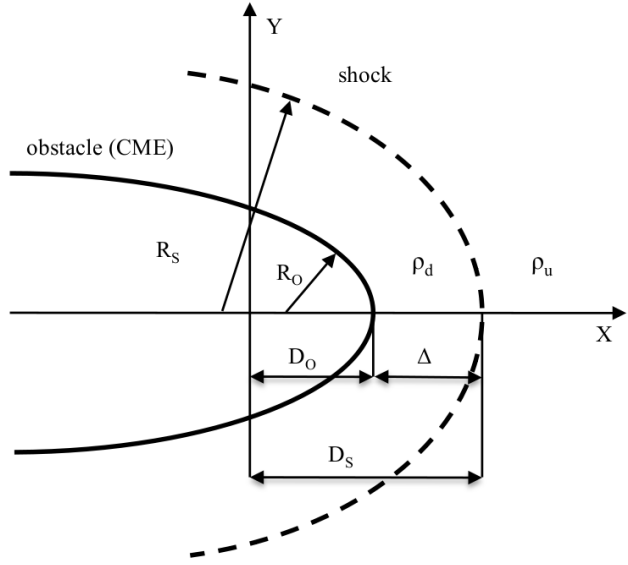}
	\caption{Diagram of the various quantities used to describe the shock and CME.}
	\label{fi}
\end{figure}

The shape, size, and standoff distance of a shock are controlled by several factors: the shape and size of the obstacle;  the velocity difference between the obstacle and the medium with respect to the sonic speed (i.e., the Mach number); and the properties of the medium, such as the ratio of specific heats ($\gamma$) and the magnetic field. Relationships between the shock standoff distance and the Mach number have been derived by a number of different authors. The well known semi-empirical relationship of \cite{Sieff:1962p19} has the form
\begin{equation}
	\label{Eq1}
	\frac{\Delta}{D_{O}}=0.78 \frac{\rho_{u}}{\rho_{d}},
\end{equation}
which was derived for a spherical object, where $\Delta$ is the shock standoff distance, $D_{O}$ is distance from the centre to the nose of the obstacle, in this case the radius,  and $\rho_{u}$, $\rho_{d}$ are the densities upstream and downstream of the shock respectively. Using gas-dynamic theory, \cite{Spreiter:1966p8946} demonstrated that $\rho_{u} / \rho_{d}$ could be written in terms of the the upstream sonic Mach number, $M_{s}$, and the ratio of specific heats $\gamma$:
\begin{eqnarray} 
	\label{Eq2}
	\frac{\Delta}{D_{O}}= 1.1\frac{(\gamma-1)M_{s}^{2}+2}{(\gamma+1)M_{s}^{2}}.
\end{eqnarray}
The increase in the coefficient in the standoff relations from 0.78 to 1.1 is due to the fact the object under consideration (Earth's magnetosphere) in Equation (\ref{Eq2}) is more blunt than a sphere; specifically, it is an elongated ellipse. Neither Equations (\ref{Eq1}) or (\ref{Eq2}) behave as expected at low Mach numbers, where the shock should move to a large standoff distance. A modification which corrects for this enables Equation (\ref{Eq2}) to be written in the form
\begin{equation}
	\label{Eq4}
	\frac{\Delta}{D_{O}}=1.1\frac{(\gamma-1)M_{s}^{2}+2}{(\gamma+1)(M_{s}^{2}-1)},
\end{equation} where the additional term in the denominator ensures the shock moves to a large distance as the Mach number approaches unity \citep{Farris:1994p9013}. They also suggested that using the obstacle radius of curvature rather than radius would be more suitable as it accounts for the shape of the obstacle, resulting in
\begin{equation}
	\label{Eq5} \frac{\Delta}{R_{O}}=0.81\frac{(\gamma - 1)M_{s}^{2}+2}{(\gamma+1)(M_{s}					^{2}-1)},
\end{equation} where $R_{O}$
is the obstacle radius of curvature.

In general, a conic section can be represented by $y(x)^{2} = 2R(D-x)+b(D-x)^2$ where $b$ is the bluntness ($b<-1$: blunt elliptic; $b = -1$: spherical; $-1 < b < 0$: elongated ecliptic; $b = 0 $: parabolic; $b > 0$: hyperbolic). The shape of the shock fronts are known to be represented by a modified conic section. One such parameterisation of the shock front, from \cite{Verigin:2003p9014}, is:
\begin{eqnarray}
	\label{Eq6}
	y^{2}(x)=2R_{S}(D_{S}-x)+\frac{(D_{S}-x)^{2}}{M_{s}^{2}-1} \nonumber \\
			\cdot (1 + \frac{b_{S}(M_{s}^{2}-1)-1}{1 + d_{S}(D_{S}-x)/R_{S}}),
\end{eqnarray} 
where $b_{S}$ is the bluntness of the shock, and $d_{S}$ is related to the asymptotic downstream slope or Mach cone (see Figure \ref{fi}).
\begin{figure*}[!ht]
	\epsscale{1.0}
	\plotone{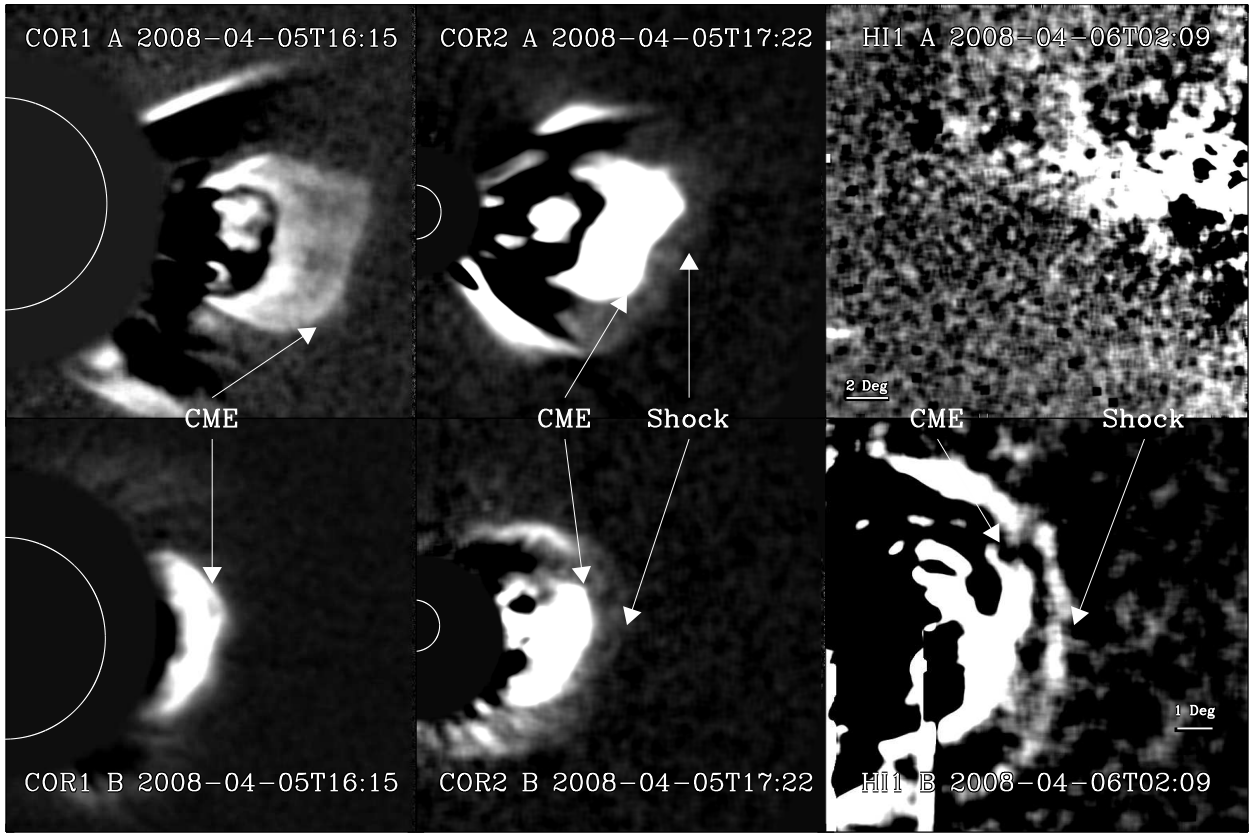}
	\caption{Simultaneous {\it STEREO} observations of the CME and shock front from \textit			{Ahead} (top row) and \textit{Behind} (bottom row). The CME and shock are indicated 			on the individual images where applicable. No CME or shock is visible in the HI 1 A 			observation (the Sun is off the right hand edge of the HI 1 A image). These images 				have been severely clipped and smoothed to make the shock more discernible. The 			shock is clearly visible in the accompanying movies (M1, M2).}
	\label{f1}
\end{figure*}

The relationships between the standoff distance and the Mach number have been investigated from a number of perspectives, including numerical modeling, analytical relations, laboratory experiments and {\it in-situ} measurements of planetary bow shocks \citep{Spreiter:1995p433, Spreiter:1980p6769}. These have shown that in general the semi-empirical relations provide an adequate description of shocks, with the low Mach regime being an exception \citep{Verigin:2003p9014}. Depending on the physical, context the sonic Mach number ($M_{S}$) can be replaced with the magnetosonic Mach number ($M_{MS}$), when dealing with plasmas such as the solar wind and CMEs. It has been shown that using gasdynamic relations when dealing magnetised plasmas works well when the MHD mach numbers are high. It also provides a good approximation when the Alfv\'{e}n ($M_{A}$) or fast magnetosonic ($M_{MS}$) Mach numbers are low and these Mach numbers are substituted for the gasdynamic ($M_{S}$) Mach numbers \citep{Fairfield:2001p9420}.

Standoff distances of CME-driven shocks have been investigated from an {\it in-situ} perspective by many authors \citep[e.g.,][]{Russell:2002p8853, Lepping:2008p125,Odstrcil:2005v110}. \cite{Russell:2002p8853} found the shock standoff distance ($\Delta$; thickness of magnetosheath)  was of the order of 21\,R$_{\odot}$ at 1\,AU. \cite{Lepping:2008p125} derived an average $\Delta$ of about 8\,R$_{\odot}$ at 1\,AU. However, when considering the CME radius (flux rope radius) as $D_{O}$ the typical $\Delta$ expected from Equation (3) is about 5\,R$_{\odot}$ at 1\,AU. \cite{Russell:2002p8853} proposed that Equation (\ref{Eq5}) may be more suited as it accounts for the fact the CME front may not be circular and that the radius of curvature at the nose is a dominant factor in determining the standoff distance. However, they found that Equation (\ref{Eq5}) did not fit the observations either and speculated this may be due to observational effect of only measuring one of the radii of curvature of the CME. The underlying structure of a CME is believed to be a flux rope, which has two characteristic curvatures; a smaller one due the curvature perpendicular to its axis (the radius when viewed as a cross-section) and the larger curvature along the axis.

In this Letter, we investigate if the shock relations above hold for a CME-driven IP shock. Specifically, we use direct observations of a CME-driven shock observed in COR2 and HI1 instruments of the Sun Earth Connection Coronal and Heliospheric Investigation suite (SECCHI, \citealt{Howard:2008p4742}) on {\it STEREO} \citep{Kaiser:2008p1663}. In Section \ref{s_obs}, we present SECCHI observations of the CME and resulting shock and describe the analysis technique. The results of our analysis are presented in Section \ref{s_res}. We discuss our results and state our conclusions in Section \ref{s_disc}

\section{Observations and Data Analysis} \label{s_obs} The CME analysed first appeared in the COR1 \citep{Thompson:2003p1587} coronagraph images from {\it STEREO B (Behind)} at 15:55\,UT on 2008 April 5. It was most likely associated with a B-class flare from NOAA active region 10987, which was just behind the west limb as viewed from Earth. Figure \ref{f1} shows the CME as it propagates out from the Sun into the different instruments' field-of-views from 8\,R$_{\odot}$ to 120\,R$_{\odot}$. The CME was visible in both {\it A (Ahead)} and {\it B} spacecraft in the inner and outer coronagraphs (COR1 and COR2; see movie M1), but was only visible in HI1 \citep{Eyles:2009p3861} from {\it STEREO B} (see movie M2). The CME propagation direction was found to be $\sim$106$^{\circ}$ west of the Sun-Earth line, the spacecraft were at a separation angle of 48$^{\circ}$ degrees (from each other). The shock is visible as a curved brightness enhancement in both the COR2 images in Figure \ref{f1} and also the HI1 {\it B} image. The accompanying movies M1 and M2 show the shock more clearly. We have made the assumption that the curved front is a shock, there are no radio or {\it in-situ} data available to corroborate this. However, due to the CME's velocity ($\sim$1000\,km\,s$^{-1}$) and the smoothness and position of the feature ahead of the CME, it can be argued that this it is a legitimate assumption \citep{Bemporad:2010p9404, Ontiveros:2009p8787}.

The observations were reduced using \mbox{{\it secchi\_prep}} from the {\sc SolarSoft} library \citep{Freeland:1998p3546}. This corrects for a number of effects such as bias, flat-field and distortions. The coronagraphs take sequences of three polarised observations, which are combined to produce total brightness images. The COR1 and COR2 observations were used to produce standard running  difference images. The HI observations were background subtracted. Modified running difference images were then created, which account for the motion of the back-ground star field \citep{Maloney:2009p6617}.

\begin{figure}[!ht]
	\epsscale{1.0}
	\plotone{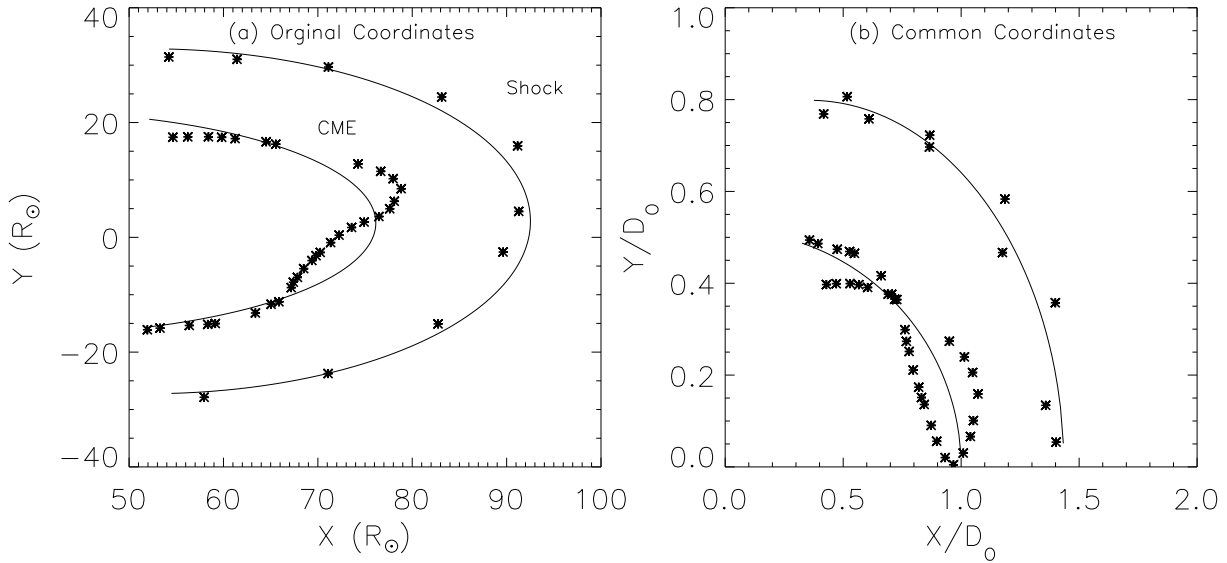}
	\caption{(a) 3D reconstruction of the CME and shock front viewed perpendicular to the 					propagation direction with the initial ellipse fits over plotted. (b) Data transformed 				into a coordinate system centred on the initial ellipse fit. Over plotted is the 						subsequent fit to the shock front using Equation (\ref{Eq6}). The data have been 				normalised with respect to $D_{O}$.}
	\label{f3}
\end{figure}

For each observation in which the CME or shock was visible the front was identified. A number of points along this front were then manually chosen. For the observations where the CME or shock was observed from both spacecraft, the front was localised in three dimensions using the tie-point method \citep{Inhester:2006p2249,2010NatCo...1E..74B,Mierla:2010p7463,Temmer:2009p5011}. As the CME or shock was only observed by one spacecraft in the HI field-of-view, we used the additional assumption of pseudo-radial propagation, based on the direction derived from COR1 and COR2 to localise the front \citep{Maloney:2009p6617}. The resulting data consisted of a series of points in 3D for the CME and shock for each observation time. Figure \ref{f3}(a) shows the 3D reconstruction of both the shock front and CME front, viewed perpendicular to the direction of propagation (assumed to be a cross-section). The techniques for deriving the 3D coordinates of features in the COR1/2 and especially the HI field-of-view are not without error. In the case of the event studied here, the CME was close ($<$10$^{\circ}$) to the plane-of-sky of STEREO A. As a result, errors in position should be small.

In order to compare with relations in Section \ref{s_intro}, the data was transformed into a coordinate system centred on the CME. To accomplish this, each CME front was fit with an ellipse. The centre coordinates of these fits were then used to collapse all the data on to a common coordinate system centred on the CME. The  shock front was fit with Equation (\ref{Eq6}), which gave the shock properties such as the shock standoff distance $\Delta$,  the Mach number $M$, and the radius of curvature at the nose of the obstacle $R_{O}$. Figure \ref{f3}(a) shows data and the initial fit, Figure \ref{f3}(b) shows the shifted data and the shock fit using Equation (\ref{Eq6}). The fast magnetosonic Mach number was calculated using $M_{ms} = ({v_{cme}-v_{sw}})/{v_{ms}}$, where $v_{cme}$ is the CME velocity, $v_{sw}$ is the solar wind velocity and $v_{ms}$ is the fast magnetosonic speed. Since $v_{sw}$ and $v_{ms}$ were not known at the position of the CME, a model corona was used to evaluate them. This was based on the Parker solar wind solution with a simple dipolar magnetic field of the form $B(r)=B_{0}(R_{\odot}/r)^{3}$, where {$B_{0}$} was 2.2\,G at the solar surface \citep{Mann:2003p9016}. For each of the paired CME and shock observations the standoff distances $\Delta$ (=$D_{S}-D_{O}$) were obtained by three different means: (i) using the 3D coordinates of the furthest point ($\textrm{max}(h)$, where $h=\sqrt{x^2+y^2+z^2}$) on the shock and the CME as $h_{shk}$ and $h_{cme}$ respectively, (ii) the previous method can be applied but to the data in the common coordinate system which gave $D_{O}$ and $D_{S}$, and (iii) the front fitting procedure also produced standoff distances. However the results of method (i) cannot be used with the relations from Section \ref{s_intro} as they are not in a CME/obstacle centred coordinate system, but the results from method (ii) and (iii) can be compared to Equations (\ref{Eq2}), (\ref{Eq4}) and (\ref{Eq5}).

\section{Results}
\label{s_res}

\begin{figure}[!ht]
	\epsscale{0.75}
	\plotone{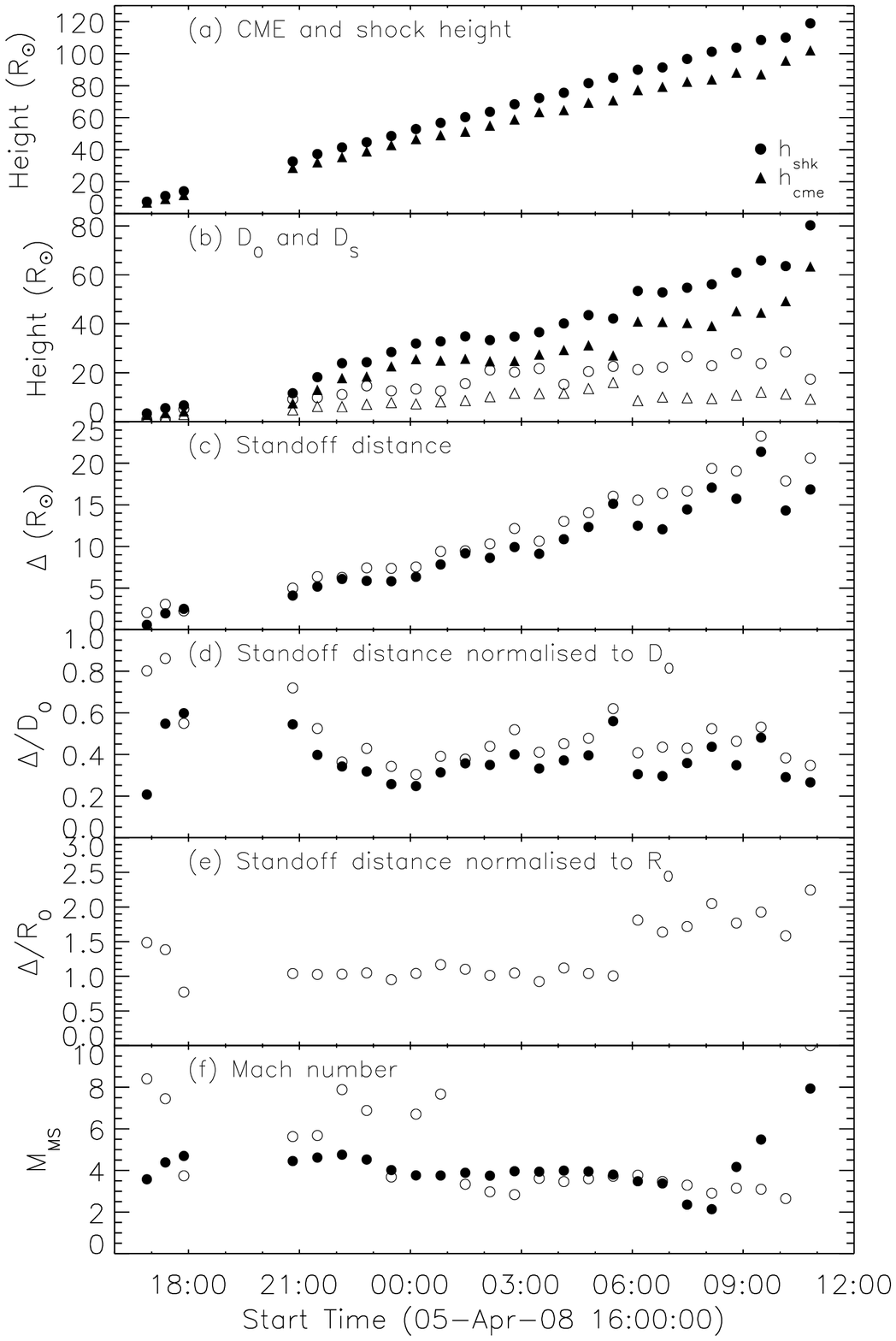}
	\caption{Shock properties derived directly from the observations (filled symbols) and from 			fits to the shock and CME (hollow symbols) as a function of time. (a) The maximum 			height of the CME front (triangles) and shock front (circles). (b) The distance to front of 		CME ($D_{O}$) and shock ($D_{S}$) in CME centred coordinate system. (c) The shock 			standoff distance $\Delta$. (d) The normalised standoff distance (${\Delta}/{D_{O}}$). 			(e) Standoff distance ($\Delta$) normalised by the radius of curvature of the CME ($R_			{O}$). (f) The Mach number ($M$) derived the CME velocity and model for the corona 			(filled circles) and from the fits to the shock front (hollow circles).}
	\label{f4} 
\end{figure}

A summary of the shock properties derived from the observations is shown in Figure \ref{f4}(a)-(f) as a function of time. With the exception of the CME ($h_{cme}$) and shock heights ($h_{shk}$), all the properties have been derived from the data collapsed on to a common coordinate system with respect to the CME. The gap between the first three data points and others is a result of both the CME and shock leaving the COR2 field-of-view and entering the HI1 field-of-view. The contrast between shock and background in the first three and last three observation is extremely low, making identification of the shock difficult. As a result, these points are not reliable, and should be neglected. Figure \ref{f4}(a) shows the derived heights of the CME and shock as they were tracked from 8\,R\,$_{\odot}$ to 120\,R\,$_{\odot}$ (0.5\,AU). Using a linear fit to $h_{shk}-h_{cme}$ (=$\Delta$) versus $h_{cme}$ (not shown), the extrapolated standoff distance at Earth was found to be $\sim$40\,R$_{\odot}$. Figure \ref{f4}(b) shows the distance to nose of the CME ($D_{O}$) and shock ($D_{S}$) front by (filled symbols), also shown are the values derived from fits to the shock and CME front (hollow symbols). The increasing offset between the two is due to their differing centres of the coordinate systems, as one is elliptic and the other is parabolic. Figure \ref{f4}(c) shows the standoff distance $\Delta$ derived using $D_{O}$ and $D_{S}$ (filled symbols) and from the fits to the fronts (hollow symbols). Both are in general agreement and show an increase with time. The standoff distance normalised using $D_{O}$ is shown in Figure \ref{f4}(d). The normalised standoff distance is roughly constant with a mean value of $0.37\pm0.09$. The standoff distance normalised to the radius of curvature at the nose of the CME ($R_{O}$) is shown in Figure \ref{f4}(e). The curvature could only be derived from the front fitting, as such only hollow data points are shown. Figure \ref{f4}(f) then shows the magnetosonic Mach number ($M_{MS}$) derived using: (i) the CME speed in conjunction with the coronal model (filled symbols) and (ii) the shock front fitted using Equation~5 (hollow symbols). The mean Mach number from the coronal model was 3.8$\pm$0.6, while a value of 4.4$\pm$1.6 was found using the front fitting method. The mean Mach number from both methods was 4.1$\pm$1.2.

\begin{figure}[!ht] 
	\plotone{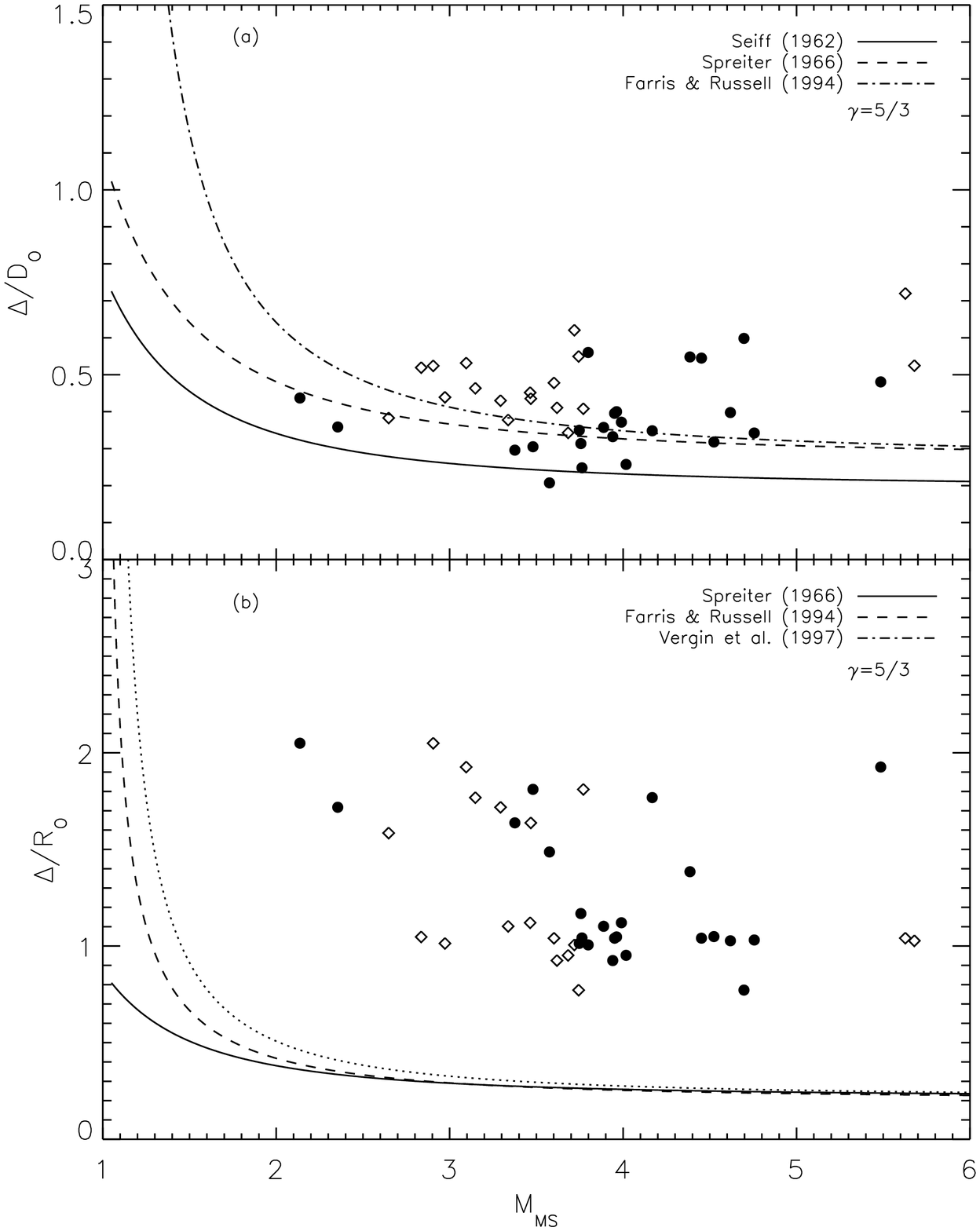}
	\caption{(a) Shock standoff distance normalised to $D_{O}$ as a function of Mach 					number. (b) Shock standoff normalised by $R_{O}$ as a function of Mach number. 				Also shown are the results of a number of semi-empirical models. Filled symbols 				indicate values derived using a coronal model, while hollow symbols indicate values 			derived using front fitting.} 
	\label{f5}
\end{figure}

Figure 5(a) shows the relationship between the normalised standoff distance ($\Delta$/$D_{O}$) and Mach number ($M_{MS}$) for a number of models. The Mach numbers were calculated using the coronal model (filled symbols) and front fitting (open symbols). The normalised standoff distances were calculated using measured values of $D_{O}$ and $D_{S}$ (filled symbols) and fits to the CME and shock fronts (hollow symbols). Both show good general agreement between our observations and the models ($<$20$\%$). The model of \citet{Sieff:1962p19} shows the poorest agreement, although this is not unexpected as it was derived for a circular obstacle and the CME is quite blunt compared to a circle. Figure 5(b) shows the relationship between the standoff distance normalised by the radius of curvature of the CME ($\Delta/R_{O}$) and Mach number for a number of models. In this case, $R_{O}$ can only be derived from the front fitting. These values are then plotted as a function of the Mach numbers derived using both methods described above (hence, each value of $\Delta/D_{O}$ appears twice). Our results do not agree with the expected relation (Equation (\ref{Eq5})) and indicate the the radius of curvature $R_{O}$ is underestimate by a factor of $\approx$3--8. One possible reason for this is that we have not considered the effect of the magnetic field of the CME and solar wind effects on the shock. However, one would expect if this had a significant effect it would also affect the other relation. It should be noted that the fast magnetosonic velocity and sonic velocity calculated from our model differ by less than 7\% after excluding the first three data points as mentioned earlier. This also suggests that the magnetic field should not play a major role. A more likely reason is due to an observational affect similar to that suggested by \cite{Russell:2002p8853}, where only one radius of curvature of the CME is observed. The observations provide a cross-sectional view of the CME along one of its axes. As a result, we have no information on the curvature along other CME axes.

\section{Discussion and Conclusions}
\label{s_disc}
For the first time, we have imaged a CME-driven shock in white light at large distances from the Sun. The shock was tracked from 8\,R$_{\odot}$ to 120\,R$_{\odot}$ (0.5\,AU) before it became too faint to be identify unambiguously. The CME was measured to have a velocity of $\sim$1000\,km\,s$^{-1}$ and a Mach number of 4.1$\pm$1.2, while the shock front stand-off distance ($\Delta$) was found to increase  linearly to $\sim$20~$R_\odot$ at 0.5~AU. The normalised standoff distance (${\Delta}/{D_{O}}$) was found to be roughly constant with a mean of 0.37$\pm$0.09\,. The normalised standoff distance derived using $D_{O}$ and $D_{S}$ and its relation to the Mach number ($M_{MS}$) were compared to previous relations and showed reasonable agreement. The normalised standoff distance (${\Delta}/{D_{O}}$) and Mach number were also derived by fitting the CME and shock front, which agreed well with theory and our other method of estimation. The fitting also allowed us to find the CME radius of curvature ($R_{O}$) enabling us to test the relationship between ${\Delta}/{R_{C}}$ and the Mach number. In this case, the derived ratios did not agree with the theoretical predictions and showed a significant deviation.

The faint nature of the shock front made its identification challenging, and thus, the front location and characterisation showed some scatter (Figure \ref{f4}). For example, the Mach numbers in Figure \ref{f4}(f) show a large amount of variability especially from the front fitting. The standoff distances in Figure \ref{f4}(c) show the same trend and the two different methods give similar results. In should be noted that the first three and last three data points show a large deviations from the rest of the data for a number of derived properties. These correspond to very low contrast observations, and hence should be ignored. The Mach number derived from our coronal model and CME position and speed, and from the shock front fitting both agree. This is a good indication that our methods accurately describe the shock even in the presence of large uncertainties.

Both sets of data for the normalised shock standoff distance $\Delta/D_{O}$ versus Mach number ($M_{MS}$) derived directly and from front fitting show good general agreement (Figure \ref{f5}(a)). The standoff distance normalised by the CME radius of curvature ($\Delta/R_{O}$) verses Mach number ($M_{MS}$) from either the fits or derived directly do not agree with any of the relations (Figure \ref{f5}(b)). Assuming that a CME can be modeled as a flux rope, it should have two radius of curvatures. Our observations are a measure of a combination of these, which depends on the orientation of the flux rope. This observational affect implies we may only be measuring the smaller of the two and this leads to the underestimation of $R_{O}$. Finally, the general agreement between the Mach number derived from our model and the Mach number derived from the fits, suggest the fitting is not the source of the problem. Using a mean Mach number of 4 to get a value of 0.26 for $\Delta/R_{O}$ ratio and the standoff distance calculated at Earth (40\,R$_{\odot}$), we can estimate the radius of curvature of this CME at Earth to be 150\,R$_{\odot}$ (0.7\,AU).

Imaging observations of CME-driven shocks opens up a new avenue for studying their fundamental properties. This type of observation will be highly complementary to radio and {\it in-situ} measurements. A complete picture of the shock could then be constructed and the derived properties from the different observations could be compared and contrasted.  Furthermore, the analysis presented here will be applicable to future observations of shocks.

\end{document}